\documentclass[sn-basic]{sn-jnl}% Basic Springer Nature Reference Style/Chemistry Reference Style
\usepackage{graphicx}
\usepackage{bm}
\usepackage{natbib}
\setcitestyle{authoryear}

\jyear{2023}%

\theoremstyle{thmstyleone}%
\theoremstyle{thmstyletwo}%

\theoremstyle{thmstylethree}%

\begin{document}

\title[On the speed of light in a vacuum in the presence of a magnetic field]{On the speed of light in a vacuum in the presence of a magnetic field}


\author*[1]{\fnm{Jonathan} \sur{Agil}}\email{jonathan.agil@lncmi.cnrs.fr}

\author[1]{\fnm{R{\'e}my} \sur{Battesti}}

\author[1]{\fnm{Carlo} \sur{Rizzo}}

\affil[1]{\orgname{Laboratoire
National des Champs Magn\'etiques Intenses (UPR 3228,
CNRS-UPS-UGA-INSA)}, \orgaddress{\street{143 avenue de Rangueil}, \city{Toulouse}, \postcode{31400}, \country{France}}}


\abstract{The nature of light, the existence of magnetism, the physical meaning of a vacuum are problems so deeply related to philosophy that they have been discussed for thousands of years. In this paper, we concentrate ourselves on a question that concerns the three of them: does light speed in a vacuum change when a magnetic field is present? The experimental answer to this fundamental question has not yet been given even if it has been stated in modern terms for more than a century. To fully understand the importance of such a question in physics, we review the main facts and concepts from the historical point of view.}

\maketitle

\section{Introduction}

In this paper, we review from the historical point of view light interactions with electromagnetic fields in a vacuum, and in particular the variation of light speed because of magnetic fields. The experimental answer to this fundamental question has not yet been given even if it has been stated in modern terms for more than a century. As we show in the following, this is not only a technical problem. Philosophers and scientists have struggled for centuries to state physics concepts in such a way that such a question could be asked.

Our subject is both very ancient and very topical with the possibility to test the Standard Model and even to go beyond the Standard Model in terrestrial laboratories and/or in astrophysical observations. 

In order to fully appreciate the importance of such a search for the influence of a magnetic field on the velocity of light, it is necessary to trace the origin of fundamental physical concepts that scientists use to describe the nature of light, magnetic fields and a vacuum. This is the goal of our paper that is somewhat arbitrarily divided in chapters concerning different ages.

The ancient times chapter treats facts from antiquity to the discovery of pneumatic vacuum by Torricelli. The modern times chapter concerns facts up to the experimental proposal by~\cite{Iacopini1979} that has been at the origin of all the existing laboratory attempts to measure this fundamental effect, and that is the natural starting point of the contemporary times chapter devoted to nowadays developments. A final section deals with conclusions and perspectives.

\section{Ancient times}

Light and magnetism have been known since prehistoric times.

Light, obviously, because it is so important in everyday life. The first treaty about it and its properties has been written by Euclid around $-300$. He summarized the fundamental knowledge of his time on subjects like light reflection, diffusion and vision into a book called \emph{Optics}~\cite{Euclid-300}, (see also~\cite{Darrigol2012} and refs. within).

Magnetism, because someone discovered the existence of a kind of stone having the strange property to attract other stones of the same nature and iron ores. It was called a magnet from Magnesia, a location somewhere in ancient Greece, apparently well known for the extraction of such a stone.

\begin{figure}
\centering
\includegraphics[width=0.4\linewidth]{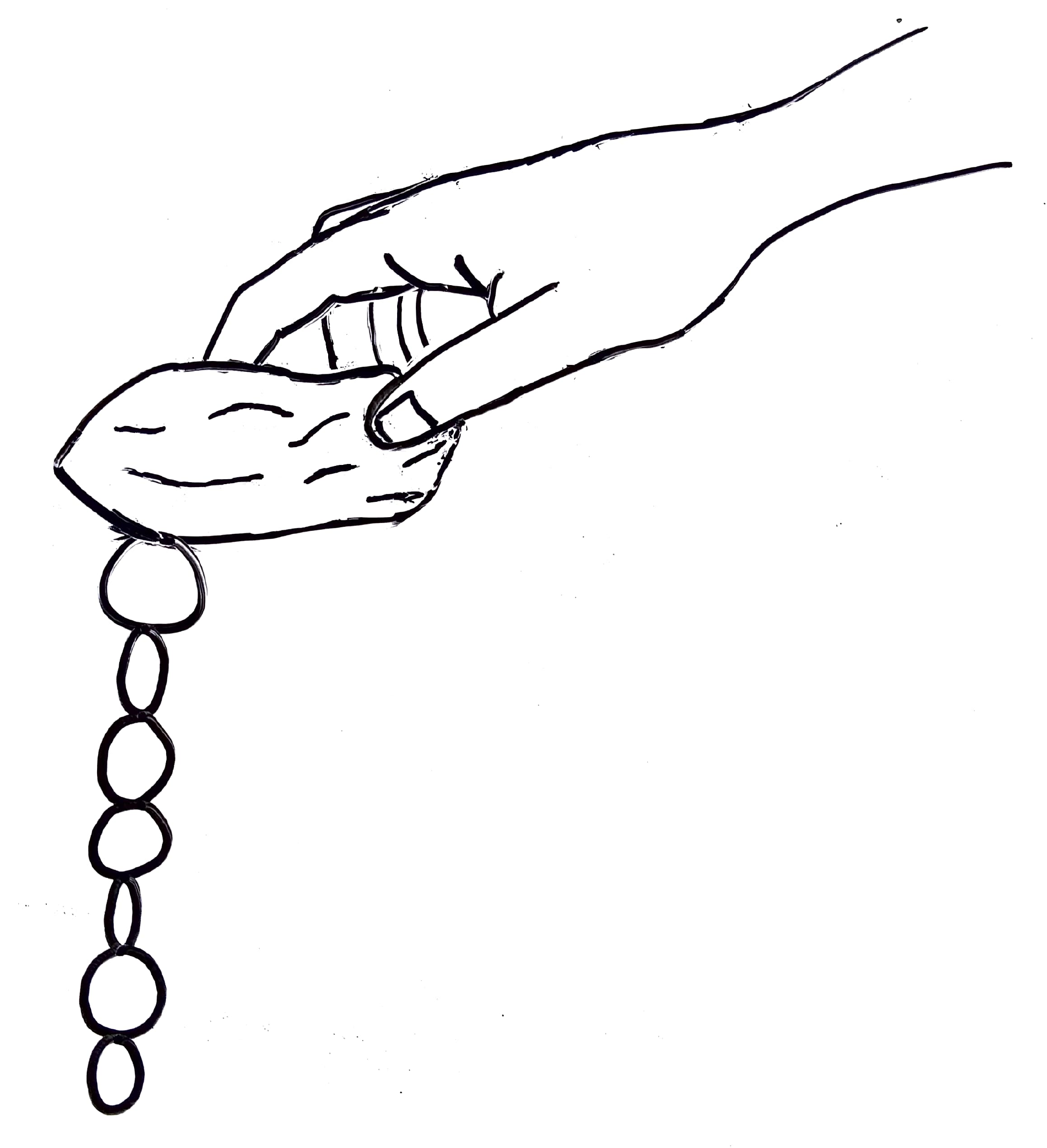}
\caption{Plato's rings experiment. A stone having magnetic properties has the ability to attract iron rings and give them the same property. However, without the stone the rings fall as they  don't attract each other anymore.}
\label{fig:platorings}
\end{figure}

One of the first written reports on a magnetic phenomenon can be found in a Plato's dialogue, \emph{Ion}, written around $-400$ : \emph{``This stone not only attracts iron rings, but also imparts to them a similar power of attracting other rings; and sometimes you may see a number of pieces of iron and rings suspended from one another so as to form quite a long chain: and all of them derive their power of suspension from the original stone''}~\cite{Plato-400}, see Fig.~\ref{fig:platorings}.

Actually, Plato was not writing about physics but about poetry arguing that \emph{``in like manner the Muse first of all inspires men herself; and from these inspired persons a chain of other persons is suspended, who take the inspiration''}~\cite{Plato-400}.

But, what are the relations between the two? Up to the 19th century the answer was straightforward: none! For example, in 1600 in the book that can be considered as the foundation of the science of magnetism, \emph{On the magnet}, William Gilbert argues that \emph{``as light comes in an instant (as the opticians teach), so much more quickly is the magnetic vigor present within the limits of its strength; and because its activity is much more subtle than light, and does not consent with a non-magnetic substance, it has no intercourse with air, water, or any non-magnetic; ... And just as light does not remain in the air above vapors and effluvia, and is not reflected from those spaces, so neither is the magnetic ray held in air or water. ... however, the magnetic power excels light, in that it is not hindered by any opaque or solid substance, but proceeds freely, and extends its forces on every side''}~\cite{Gilbert1600}. As far as we understand, for Gilbert, since light is a non-magnetic substance, it has no intercourse with the magnetic power, so light and magnetic power permeates the whole space not interfering with each other. 

In these few lines Gilbert also gives, as a well known fact, that light and magnetic power come in an instant, \emph{i.e.} in nowadays terms, that light velocity is infinite. This was considered so evident that Descartes wrote in a letter to Isaac Beeckman in 1634 that \emph{``The instantaneous propagation of light is to me so certain that if its falsity could be shown, I would be ready to admit my complete ignorance of Philosophy.''}~\cite{Descartes1634a},~\cite{Darrigol2012}.

\begin{figure}
\centering
\includegraphics[width=0.3\linewidth]{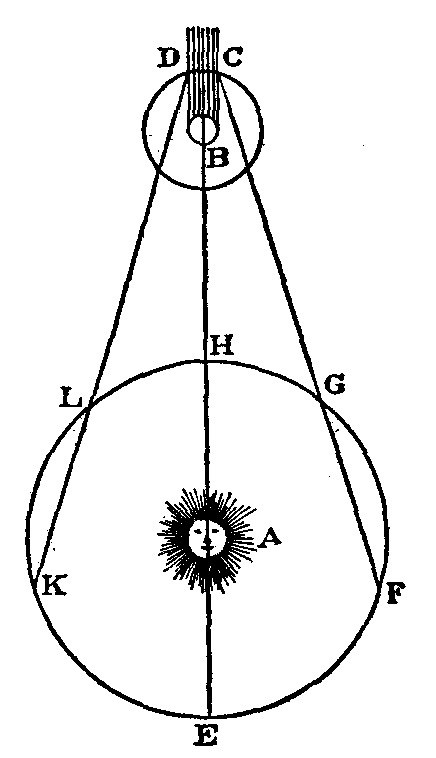}
\caption{In the 17th century, Ole R{\o}mer observed a variation of the time between two eclipses of Io, a satellite of Jupiter, depending on the position of the Earth in its orbit. Because of the change of the distance from Earth to Jupiter, he concluded that light has a finite velocity. The image is from a reproduction of \cite{Roemer1676,1677a} in \cite{Magie1935}.}
\label{fig:roemer}
\end{figure}

First measurement of a finite light velocity dates back to 1676. It was performed by R{\o}mer by observing the eclipses of Io the satellite of Jupiter at different positions of the Earth in its solar orbit~\cite{Roemer1676,1677a}, as shown in Fig.~\ref{fig:roemer}. This observation was reported in~\cite{1677a}: \emph{``The necessity of this new Equation of the retardment of Light, is established by all the observations that have been made in the R.Academy, and in the Observatory for the space of eight years, and it hath been lately confirmed by the Emersion of the first Satellite observed at Paris the 9th of November last at 5 a Clock, 35'. 45". at Night, 10 minutes later than it was to be expected, by deducing it from those that had been observed in the Month of August, when the Earth was much nearer to Jupiter: Which M.Romer had predicted to the said Academy from the beginning of September.''}

R{\o}mer was at that time in Paris as the assistant of Cassini and Picard, two astronomers members of the ``Académie des Sciences''. Cassini, in particular, had published a detailed table of the movements of Jupiter's satellites. Its goal was to use their observations to determine the value of the longitude of a point on the Earth's surface. The idea, coming from Galileo himself, the discoverer of Jupiter's satellites, was to use Jupiter as a sort of master clock. The longitude being worked out by comparing the satellite observation in Paris, for example, and any other spot on Earth. The determination of the longitude was a very serious problem to be solved while European ships were busy with exploration journeys~\cite{Cohen1940}.

At the end of the 17th century, the finiteness of the velocity of light was still questioned~\cite{Darrigol2012}. The final confirmation came in 1729 by Bradley, who observed the stellar aberration, \emph{i.e.} the motion of the apparent position of a star, that could be explained only by a combined effect of the motion of the Earth and a finite velocity of light~\cite{Bradley1728}.

Actually, for centuries the title of this paper would have been simply considered as nonsense. Not only because we talk about light velocity but also because we assume the existence of a vacuum, a thing that has been considered impossible for centuries following the Aristotle teaching. Around $-350$ Aristotle considered that vacuum is a fundamental concept in science and he wrote in his treaty called \emph{Physics} that \emph{``the investigation of questions about the vacuum must be held to belong to the physicist - namely whatever it exists or not, and how it exists or what it is''}~\cite{Aristotle-350} but he concluded that \emph{``Since we have determined the nature of place, and void must, if it exists, be place deprived of body, and we have stated both in what sense place exists and in what sense it does not, it is plain that on this showing void does not exist''}~\cite{Aristotle-350}. Aristotle had the last word for almost 2~000 years.

It is worth mentioning that Aristotle also reported an observation proving that a vacuum cannot exist concerning the motion of bodies: \emph{``Further, in point of fact things that are thrown move though that which gave them their impulse is not touching them, either by reason of mutual replacement, as some maintain, or because the air that has been pushed pushes them with a movement quicker than the natural locomotion of the projectile wherewith it moves to its proper place. But in a void none of these things can take place, nor can anything be moved save as that which is carried is moved. ''}~\cite{Aristotle-350}. % [Physics IV ($\Delta$), 8, 215 a-b].
In other words, all the space has to be filled by some substance otherwise no motion is possible. It is thanks to the action of this supporting medium that bodies can continue their motion once they are separated by the source of their impulse.

\begin{figure}
\centering
\includegraphics[width=0.3\linewidth]{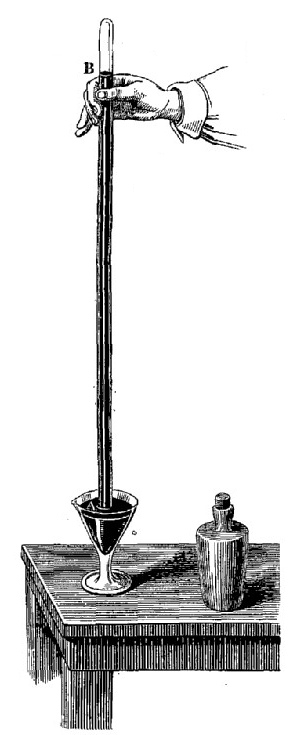}
\caption{Torricelli's experiment. A tube filled with mercury is turned upside down inside a container filled of the same liquid. One observes that the level of the mercury in the tube decreases and stops at a fixed height. The space at the top of the glass tube is no more filled with mercury. Is it a vacuum? Illustration extracted from \cite{Ganot1859}.}
\label{fig:Torricelli}
\end{figure}

Vacuum became a physical object only in 1644 thanks to Torricelli and his famous experiment~\cite{Torricelli1644}, illustrated in Fig.~\ref{fig:Torricelli}, and repeated extensively and very fruitfully by Pascal~\cite{PascalOeuvre}.

The goal of Torricelli was \emph{``not to create simply a vacuum, but to built an instrument to show the changes of air, sometimes heavier and thicker, sometimes lighter and subtler''}~\cite{Torricelli1644}. While the fact that Torricelli's experiment provides an evidence of the weight of air was accepted worldwide rapidly, the existence of a vacuum was immediately questioned. In particular, Descartes' vision of the world was that the whole cosmos was filled by a subtle matter, namely the ether, that was able to pass through the pores of the glass constituting Torricelli's tube preventing a vacuum to exist~\cite{Cartesio}. This is what Descartes gave as an explanation to Pascal himself in a meeting in 1647 as reported by Jacqueline Pascal, his sister : \emph{``Then we got to the vacuum, and Mr. Descartes in a very serious manner, when he was told about} [Torricelli's] \emph{experiment and asked what he thought entered the tube, said that it was his subtle matter.''}~\cite{Blaiseletter}. Actually, Descartes has filled the cosmos with his subtle matter also to allow the propagation of light, as Aristotle did to justify body motion. Now, the fact that the ether is needed for light propagation \emph{``will give the possibility to those who believe in the existence of a vacuum to note that there is none of it in the upper part of} [Torricelli's] \emph{tube, and that the space that the mercury has left is filled by some matter since the visible objects which are on the other side still act on our eyes to give us the same sensation as before; what they could not do if there was a vacuum because their action would be stopped; and even when one's eye is against the tube, one should not see more than in darkness, or as if there was an opaque body in front of it; which does not happen.''}~\cite{Rohault1671}. This is what can be read in a 17th century treaty on physics written by Rohault, one of Descartes' followers.

The concept of ether comes from ancient times but before Descartes it had no mechanical properties. Descartes can be considered as the first to bring the ether into modern sciences~\cite{Whittaker1989}.

\section{Modern times}

But, what is the nature of the substance filling all the cosmos? What are its properties? Does it move? How does it interact with the other material bodies?

From the point of view of classical philosophy, a vacuum is a much simpler object to define than the ether. A vacuum is literally nothing, no properties has to be assigned to it.

From the point of view of Newtonian gravitation, since we do not observe any friction caused by the ether on celestial bodies, a vacuum or a free ether, \emph{i.e.} an unperturbed ether is the same from the phenomenological point of view~\cite{Whittaker1989}. 

Things get more complicated when light comes into play. Centuries of theories and experiments have been necessary to explore all the aspects concerning the ether~\cite{Whittaker1989}. As far as light is concerned, it was assumed to be a vibration of the ether itself just as sound is a vibration of a material body~\cite{Darrigol2012}.

Moreover, in 1845, Faraday observed that the linear polarisation of light rotates when light propagates in a medium in the presence of a magnetic field parallel to its direction of propagation~\cite{Faraday1846_1,Faraday1846_2}. This was considered a clear indication of the connection between light and magnetism~\cite{Whittaker1989}. As a matter of fact, Faraday was convinced \emph{``that the various forms under which the forces of matter are made manifest have one common origin; or, in other words, are so directly related and mutually dependent, that they are convertible, as it were, one into another, and possess equivalents of power in their action.''} (see~\cite{Darrigol2003} and references within). The discovery of what we call nowadays the Faraday effect consisted for him \emph{``in illuminating a magnetic curve or line of force and in magnetising a ray of light.''} (see~\cite{Darrigol2003} and references within).

Maxwell in his 1873 book \emph{``A treatise on electricity and magnetism''} explains light propagation like this: \emph{``According the theory of emission, the transmission of energy is affected by the actual transference of light-corpuscules from luminous to the illuminated body, carrying with them their kinetic energy, together with any other kind of energy of which they may be receptacles. According to the theory of undulation, there is a material medium which fills the space between the two bodies, and it is by the action of contiguous parts of this medium that the energy is passed on from one portion to the next, till it reaches the illuminated body. The luminiferous medium is therefore, during the passage of light through it, a receptacle of energy. In the undulatory theory, as developed by Huygens, Fresnel, Young, Green, $\&$c., this energy is supposed to be partly potential and partly kinetic. The potential energy is supposed to be due to the vibratory motion of the medium. We must therefore regard the medium as having a finite density''}~\cite{Maxwell}. 

No ether, no electromagnetic waves, but what about electricity and magnetism? They can be explained by translation or rotation of the ether itself~\cite{Whittaker1989}. Now, if an electric or a magnetic field perturbs the free ether, and some light propagates into it, the light celerity has to be affected as well and change.

As far as we know, the most ancient experimental attempt to see a variation of the index of refraction of light, \emph{i.e.} of light velocity, when light propagates very close to bodies charged with electricity has been reported by H. Wild in 1864~\cite{Wild1865}. The author explains his motivation for publishing his work as follows: \emph{``Recently, several physicists have directly or indirectly supposed that the light ether and the electric fluid are the same thing following a unitary vision of the nature of the electricity. One immediate consequence is that the density of the ether in a body positively electrified has to be bigger or smaller than in a body negatively electrified. Since November 1860, I performed some experiences to test experimentally the consequences of this hypothesis. Since all of them has given a negative result, and that for some of them I have employed somewhat inadequate apparatus, I had avoided to publish them. Since then, discussion with some physicists, friends of mine, have shown me that others have also tested experimentally this hypothesis with as few success as me. I have thought then that it was worth while to report to the public my investigations on this point, that I have been somewhat finishing this autumn, even if it has done nothing more than confirming my previous negative results.''}

A few years later in 1873, A. Roiti tested experimentally another hypothesis \emph{i.e.} that an electric current is an ether current~\cite{Roiti1873a,Roiti1873}. If it is so, some light propagating in the direction of the electric current will have its velocity increased because of the sum of its velocity with the ether one. No variation of the velocity of light was observed, so that the ether velocity was evaluated to be smaller than 200~m/s which was much less than the velocity of propagation of electricity in a circuit. 

Let's note also that in 1884 Lecher reported~\cite{Lecher1884} a repetition of Roiti's experiment, always with a negative result, and also a search for a current induced in a solenoid by the passage of a circularly polarised light. The motivation was \emph{``to know if the rotation of the polarization plan of the light by an electric current is a reversible phenomenon or not''}. The result was negative as well.

\begin{figure}
\centering
\includegraphics[width=0.9\linewidth]{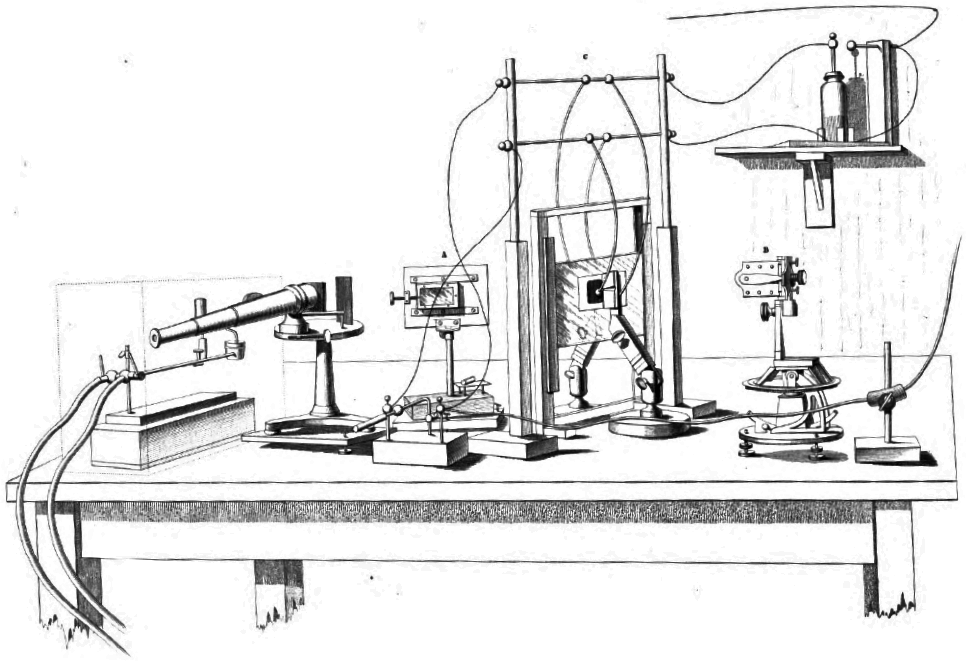}
\caption{Wilberforce apparatus used to improve on Rioti's results. The interferometric device is composed of a thick mirror (A on the illustration) and two right angled mirrors (B), in between a displacement current can be applied to a dielectric (C). Interference fringes were observed thanks to a telescope. Extracted from~\cite{Wilberforce1887}.}
\label{fig:Wilberforce}
\end{figure}

A more precise result than the Rioti's one, 41~m/s, was obtained by Wilberforce in 1887~\cite{Wilberforce1887} using a more advanced interferometric technique and a condenser to create a displacement current in a dielectric in which light propagated, the experimental apparatus is illustrated in Fig.~\ref{fig:Wilberforce}. As the author states : \emph{``according to the view of electromagnetic field taken by Maxwell, there is a certain ether or medium pervading the whole space, with which the molecules of ordinary matter are in the same way associated or connected, and which is the seat of all electric and electromagnetic forces, ... we must assume at the same time that if any portion of the conducting matter is set in motion the medium associated with it will move with the same velocity.''} The result being negative, this was not the case.

The effect of a magnetic field and again an electric field has also been tested by Lodge as reported in 1897~\cite{Lodge1897}. These are the author's conclusions \emph{``without further delay I conclude that neither an electric nor a magnetic transverse field confers viscosity upon the ether, nor enables moving matter to grip and move it rotationally.''}

All these experiments were performed using interferometric techniques and, in a sense, their goal was actually the detection of the ether movement corresponding to the electromagnetic phenomena more than the observation of a variation of the light celerity which was considered as a straightforward consequence of the ether motion. 

On the other hand, a few years before, the subject of the present paper has been discussed in an official meeting of physicists with a more modern attitude without mentioning the ether at all. Actually, in 1889 at the Toronto Meeting of the American Association for the Advancement of Science, H.T. Eddy presented theoretical results concerning a possible variation of the speed of light in the presence of a magnetic field. \emph{``In the discussion that followed \ldots{} Prof.~Morley suggested a form of apparatus which would detect the suspected change of velocity \ldots{} The whole matter was of such interest that the section of the Association \ldots{} obtained from the research funds \ldots{} a grant of money with which to construct the apparatus and make the experiment. \ldots{} The optical part of the apparatus consists of the interferential refractometer used by Michelson in his experiments to determine whether light moves with the same velocity in different directions in the solar system.''} With these words Morley tells us the origin of the experiment which he carried out between 1889 and 1898~\cite{Morley1898}. 

What has happened in the second half of the 19th century to justify that the ether concept could be given up? The ether could not stand the results of the experiments measuring the light velocity when the relative motion of matter with respect to the ether has to be taken into account.

Physicists knew that light velocity in a material medium is $c/n$, where $n$ is the index of refraction which depends on the medium. Now, what about if the medium is moving at a velocity $v_m$? If the ether does not move with it light should keep on propagating at $c/n$, if the ether moves with the medium $c/n$ and $v_m$ have to sum up.

\begin{figure}
\centering
\includegraphics[width=0.9\linewidth]{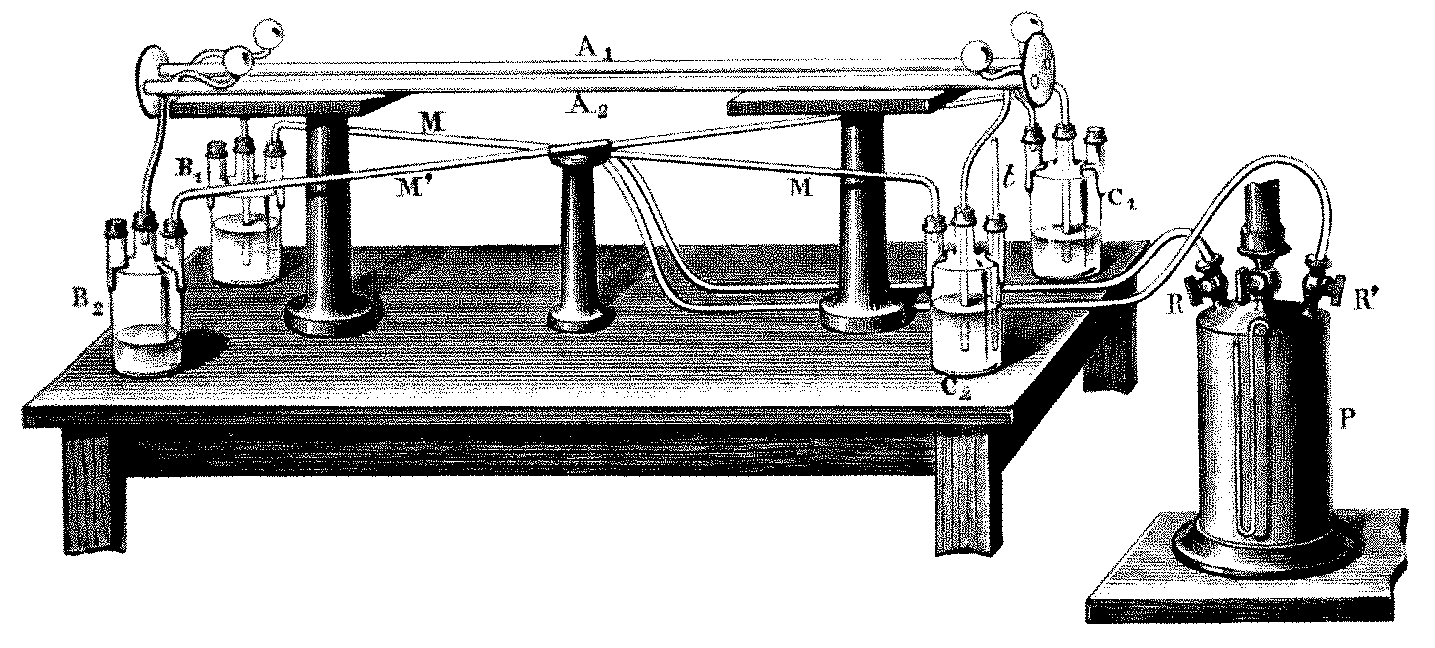}
\caption{Fizeau's apparatus used to measure light velocity in a moving fluid. Only the liquid system is represented here, allowing the fluid to flow circularly in the tubes A$_1$ and A$_2$. A mirror is used to make light circulate in both tubes, light going around in one way interfere with the one going the other way. Fringes displacement allow to calculate light velocity in function of the fluid flow. Extracted from \cite{Mascart1889}.}
\label{fig:Fizeau}
\end{figure}

To find the correct answer, Fizeau performed an experiment in 1851 whose result was puzzling~\cite{Fizeau1851}, a part of the apparatus is represented in Fig.~\ref{fig:Fizeau}. Fizeau's results suggested that the ether was ``partially'' dragged since the light velocity $v_l$ was intermediate between the two expected values. If the light propagation is parallel to the body velocity $v_l \approx c/n + v_m\left(1-1/n^2\right)$.

\begin{figure}
\centering
\includegraphics[width=0.6\linewidth]{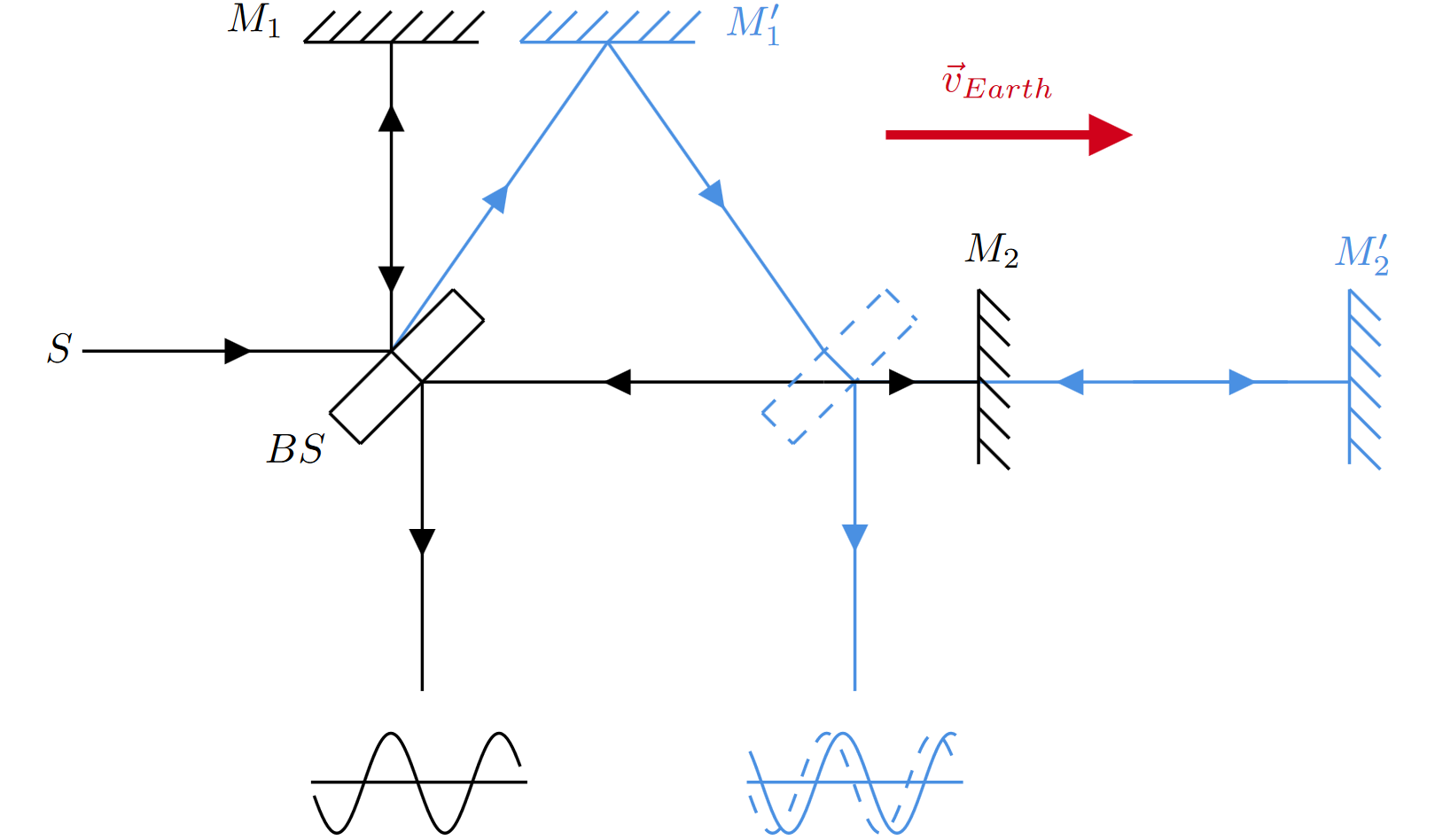}
\caption{The Michelson-Morley experiment: a beamsplitter is used to divide and recombine a light beam along two orthogonal direction. If an ``ether wind'' due to the velocity of Earth with respect to the ether exists it would cause an observable fringe shift.}
\label{fig:michelson}
\end{figure}

Well, one could even accept this but in 1887 Michelson and Morley published their null result obtained when looking for the ether drift, their famous interferometer is represented in Fig.~\ref{fig:michelson}. Light velocity measured in a moving frame is always $c$ regardless of the relative motion with respect to the ether, \emph{i.e.} the ether moves with the body. This is not compatible with the Fizeau's experimental result, and finally Einstein's special relativity turned the ether concept in a superfluous one.

In their book titled \emph{The evolution of Physics}, Einstein and Infeld state their point of view~\cite{EinsteinInfeld}: \emph{``All assumptions concerning ether led nowhere! The experimental verdict was always negative. Looking back over the development of physics we see that the ether, soon after its birth, became the ``enfant terrible'' of the family of physical substances. First, the construction of a simple mechanical picture of the ether proved to be impossible and was discarded. This caused, to a great extent, the breakdown of the mechanical point of view. Second, we had to give up hope that through the presence of the ether one coordinate system would be distinguished and lead to the recognition of absolute, and not only relative, motion. This would have been the only way, besides carrying the waves, in which ether could mark and justify its existence. All our attempts to make the ether real failed.''}

\begin{figure}
\centering
\includegraphics[width=0.6\linewidth]{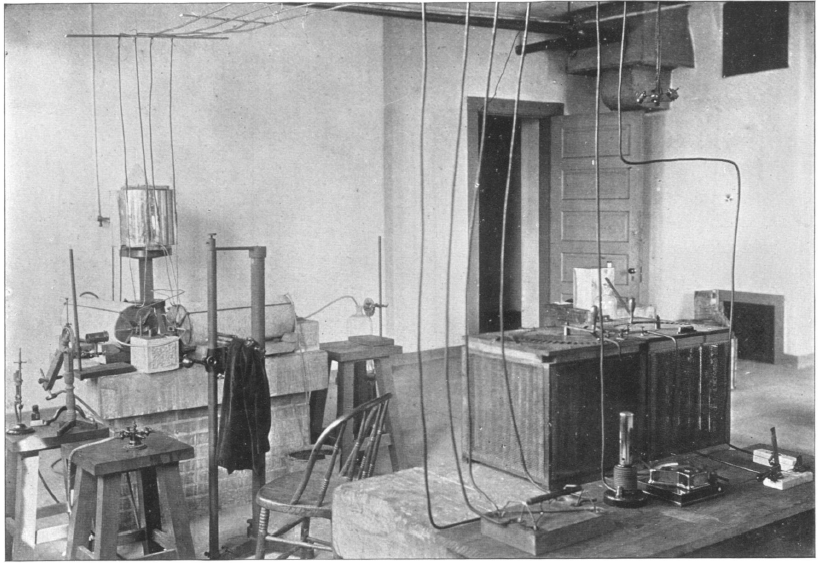}
\caption{The experiment of Eddy, Morley and Miller, they describe their apparatus in~\cite{Eddy1898}: \emph{``At the extreme right are seen the commutator, amperemeter, and resistance coils, used in managing the electric current. The wooden stand at the extreme left carries the source of light and the condensing lens. The adjacent stone pier carries the coils; between them is seen the cubical block of stone which supports the diagonally placed mirrors \ldots{} Apparently just above this block, but really some yards to the rear, is a double tank supplying water and carbon bisulphide to the apparatus. The reading telescope with which the observations were made is marked by the hanging cloth. On the left edge of the pier is seen an iron stand carrying a Nicol's prism for polarizing the ray of light sent through the apparatus; on the wooden stand to the right of the pier is seen the analyzer, by means of which the rotation produced was measured while the current was adjusted so as to secure the rotation desired.''}. Extracted from~\cite{Eddy1898}.}
\label{fig:morley}
\end{figure}

Let's come back to Morley's experiment. A Physical Review paper~\cite{Eddy1898}, also published in 1898, gives more explanations on the theoretical motivations. The goal was to test Eddy's prediction that a circularly polarised light has a velocity of propagation different from the one of the linearly polarised light. A magnetic field of about 0.2~T was present in one of the arms of the interferometer to rotate the polarisation thanks to the Faraday effect in a cell filled with carbon bisulphide. A similar cell was present in the other arm of the interferometer but without magnetic field; Fig.~\ref{fig:morley} is a photograph of their apparatus.

The results were again negative, they write: \emph{``The result reported at the Boston meeting of the American Association is, that we are confident that when light corresponding to the solar D line is passed through one hundred and twenty centimetres of carbon bisulphide in a magnetic field which produces rotation by half a circumference in the plane of its polarisation, there is no such change of velocity as one part in sixty million, and probably no such change in a hundred million.''}~\cite{Eddy1898}.

With today's knowledge, the theoretical foundations are not very clear. Actually, the magnetic field is only indirectly the cause of the hypothetical effect that, as far as we understand, should also exist in optical active media like sugar solutions where a rotation of polarisation is observed without the need of a magnetic field.

Let's finally note that Miller, one of the two experimentalists, tried for several years to repeat the Michelson-Morley experiment and he reported in 1925 a positive result being at the origin of a revival of the ether drift experiments~\cite{Navarro2018} that ultimately confirmed a definitive negative result.

\begin{figure}
\centering
\includegraphics[width=0.8\linewidth]{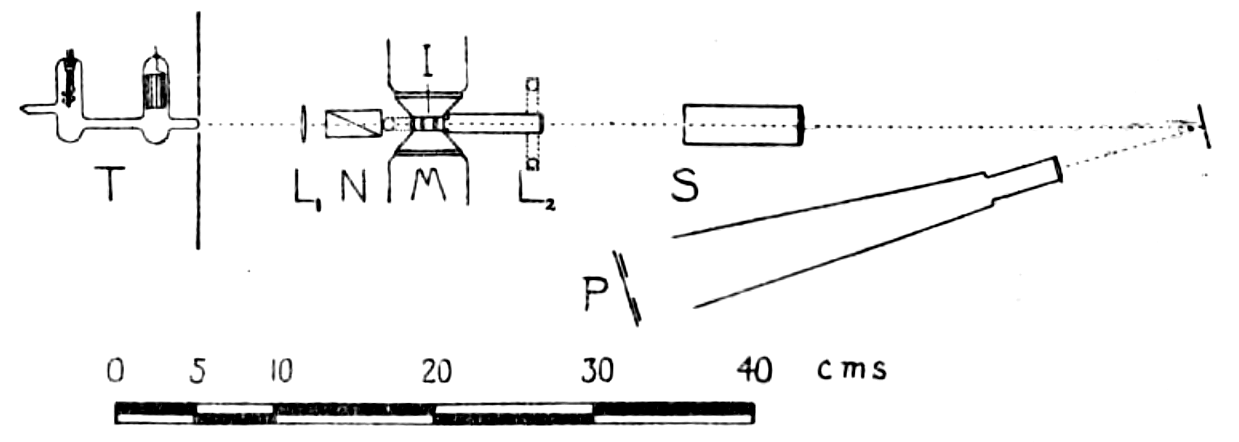}
\caption{Watson's experiment to search an influence of a transverse magnetic field on the speed of light in vacuum.  He used a Neon tube discharge as a photon source, T, that he polarised using a Nicol prism, N, then used a Fabry-Perot interferometer, I, inside a magnet, M, to create an interference pattern. He expected fringe broadening but measured none outside of uncertainties. Extracted from~\cite{Watson1929}.}
\label{fig:Watson}
\end{figure}

It is only in 1929 that a new attempt to observe the effect of a magnetic field on light is reported, when Watson published in the Proceedings of the Royal Society of London the results of his experiment carried out at the Cavendish Laboratory~\cite{Watson1929}. The thirty years that have passed by since the last attempt by Morley have seen the birth of the theory of Relativity and Quantum Mechanics. Watson's theoretical motivation takes this into account: \emph{``In modern physics there have developed two complementary -- and apparently mutually contradictory -- modes of description of radiation processes and of the motion of molecules, atoms, electrons and protons. \ldots{} The development of the analogy between photons and entities of the second class (atoms, electrons, etc.) has already reached the stage where it is possible to give a wave description of the motion of the particles in all cases and to assign to the particles an energy-momentum four-vector within the limits of the Heisenberg's principle. There remains, however, a fundamental distinction in current theory. All entities in the second class have electromagnetic particle properties while none have been assigned to the photon. \ldots{} The present investigation was carried out with the object of detecting, if possible, the existence of the magnetic moment of a photon''}. Watson's experiment is represented in Fig.~\ref{fig:Watson}.

The light, polarised by a Nicol prism, passes through a Fabry-Perot interferometer and the interference fringes are observed. Inside the Fabry-Perot a magnetic field $\mathbf{B}$ ($B \approx 1$~T) perpendicular to the direction of propagation is present; if the photon has a magnetic moment $\bm{\mu}$, the interaction energy $\Delta E$ is necessarily $\Delta E=-\bm{\mu}\cdot\mathbf{B}$, which also corresponds to a change in frequency of amplitude $\Delta\nu/\nu=\bm{\mu}\cdot\mathbf{B}/h\nu$. By turning on the magnetic field, one should see the interference fringes move, if an effect exists. Watson's result was zero, the variation in the refractive index $\Delta n$ was less than $4\times 10^{-7}$. Let's note that according to the nowadays Standard Model, a zero magnetic moment of the photon is expected~\cite{Workman2022}.

Apparently, the experiment has been known around. It is quite amusing to learn by Kapitza's memories~\cite{Kapitsa1980} that: \emph{``In the 1930s, in Cavendish's laboratory, I developed a method of obtaining magnetic fields one order stronger than had previously been attained.}~[See \cite{Kapitza1927}] \emph{In a conversation Einstein tried to persuade me to study experimentally the influence of a magnetic field upon the velocity of light. Such experiments had been conducted, and no effect was discovered. In my magnetic fields it was possible to raise the limit of accuracy of measurement by two orders of magnitude, because the effect was dependent on the square of the intensity of the magnetic field. I protested to Einstein that according to the existing picture of electromagnetic phenomena, I could not see from whence such a measurable phenomenon would come. Having found it impossible to prove the need for such experiments, Einstein finally said, ``I think that} die liebe Gott \emph{could not have created the world in such a fashion that a magnetic field would be unable to influence the velocity of light.'' Of course, it is hard to counter that kind of argument''.} Very likely, Einstein visited Kapitza's laboratory in 1932 when he spent some time in Cambridge to talk with Eddington and give a Rouse Ball Lecture~\cite{Fox2018}. Watson was no more there having moved to Canada to the McGill University in 1931~\cite{QueenUniv}.

Let's note that Watson was looking for a $\mathbf{B}$ effect, and the first citation of a $\mathbf{B}^2$ effect in a vacuum, dates only from 1961~\cite{Erber1961}, but Albert Einstein seems to have understood before anybody else that the effect of the magnetic field on light velocity must depend on $\mathbf{B}^2$. A likely explanation is that, since Einstein was convinced that {\textit{c}} cannot be exceeded, a magnetic field can only reduce it \emph{i.e.} the index of refraction in the presence of a magnetic field in a vacuum has to be greater than 1 whatever is the sign of $\mathbf{B}$. $\mathbf{B}^2$ therefore, not $\mathbf{B}$.

\begin{figure}
\centering
\includegraphics[width=0.6\linewidth]{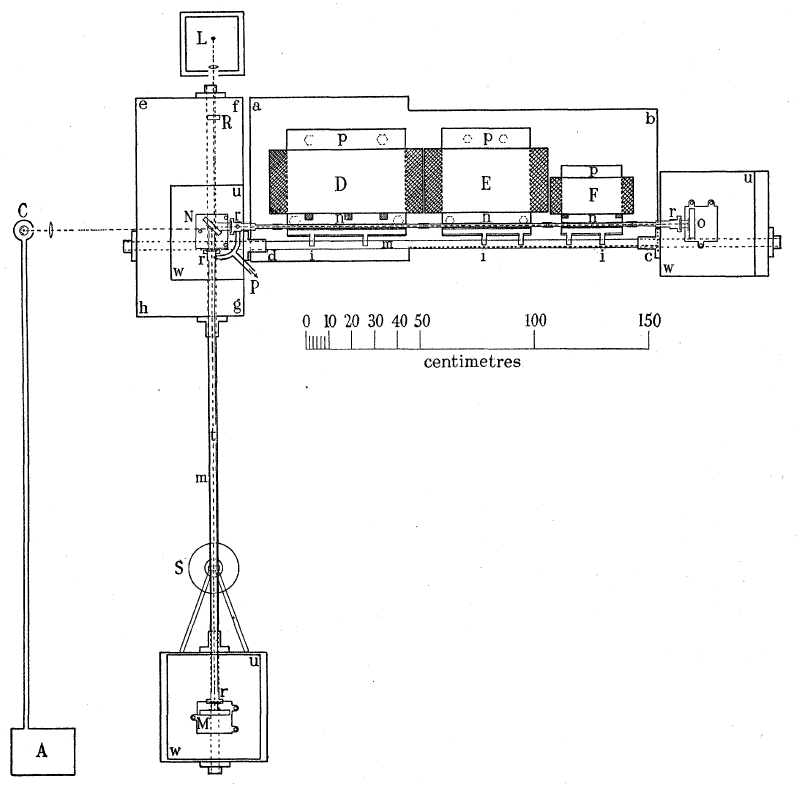}
\caption{The experiment of Farr and Banwell searched for a change of the speed of light under a transverse magnetic field in vacuum. Experimentalists used a Michelson-Morley setup where on one arm a magnetic coil capable of producing a field near 2~T was placed. If the magnetic field has an influence on the speed of light a displacement of  the interference fringes was to be observed. The result found that the change of velocity of light in vacuum is less than 14 meters per second. In the figure, the interferometer is formed of L the light source, N the beam splitter, and M and O the mirrors. The electromagnets are D, E and F, observations are made at C. Figure extracted from~\cite{Banwell1940}.}
\label{fig:farr banwell}
\end{figure}

In the same period Farr and Banwell were working on the same phenomenon in New Zealand~\cite{Farr1932}, the theoretical motivation was similar to the Watson's one. The results have been published in 1932: \emph{``In the present work, which was in progress at the time when Watson's paper was published, a different experimental method is adopted and the apparatus developed carries the same conclusions he obtained, but twenty times the sensitiveness which he ascribed to his results''}. A later article~\cite{Banwell1940}, published in 1940, presents an improved device in terms of sensitivity and therefore a result that is still zero but 30 times more precise: $\Delta n<2\times 10^{-9}$. Farr and Banwell's apparatus was a Michelson interferometer where one of the two arms was subjected to a magnetic field perpendicular to the direction of light propagation of about $2$~T, see Fig.~\ref{fig:farr banwell}.

Between the first and the second article by Farr and Banwell, everything has changed on a theoretical level. Dirac~\cite{Dirac1934}, Euler~\cite{Euler1935}, Heisenberg~\cite{Heisenberg1936}, Weisskopf~\cite{Weisskopf1936} developed the quantum theory of vacuum, that is to say worked out a theoretical relation which connects, in a vacuum, the vector $\mathbf{D}$, $\mathbf{H}$ to the vectors $\mathbf{E}$, $\mathbf{B}$ present in the equations of Maxwell in the framework of the novel quantum theory.

\begin{figure}
\centering
\includegraphics[width=0.6\linewidth]{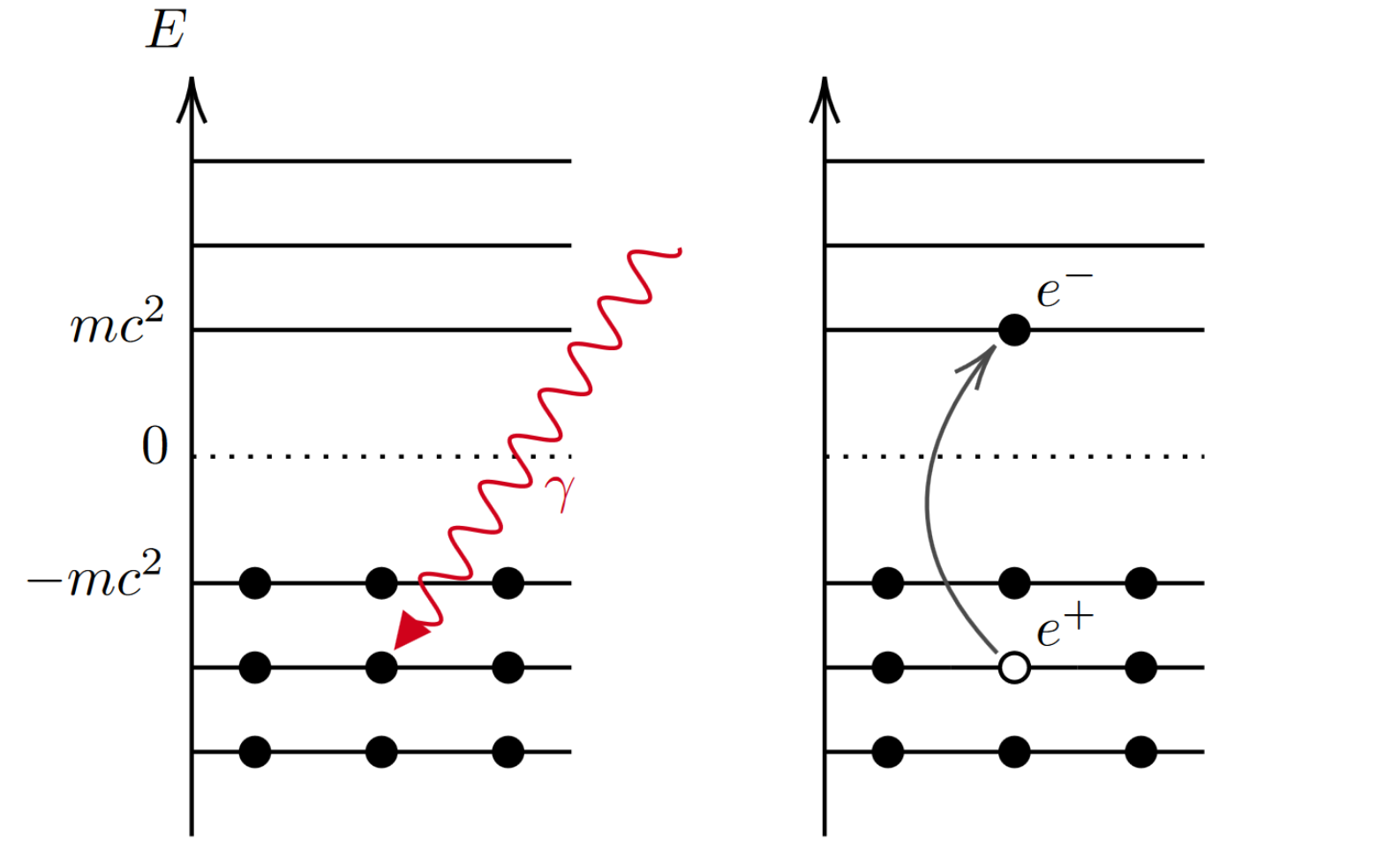}
\caption{Dirac's interpretation of the negative energy states that his equation yields. An infinite number of electrons occupy these negative energy states, a virtual electron can be created by excitation of one of these states. The virtual hole left behind is a new soon-to-be discovered particle of the same mass but of opposite charge, the positron.}
\label{fig:Diracsea}
\end{figure}

The 1936 Weisskopf's article~\cite{Weisskopf1936}, entitled \emph{The electrodynamics of the vacuum based on the quantum theory of the electron} is very instructive. The mass-energy equivalence plays a very important role: a quantum of light can be absorbed in a vacuum and transformed into matter in the presence of other electromagnetic fields thanks to the creation of an electron-positron pair. The phenomenon of absorption of light in a vacuum is incompatible with classical electrodynamics. Classically several fields can be superimposed without interacting. This is how Weisskopf sums up the phenomenon in question: \emph{``When light passes through electromagnetic fields, it will behave as if the vacuum were given a dielectric constant different from that of the unit under the influence of the fields''}. The effect of a magnetic field $\mathbf{B}$ on the propagation of light in a vacuum is therefore to make $\varepsilon$ and $\mu$, and therefore $v$ and $n$, functions of $\mathbf{B}$. To be able to calculate the new expression for the energy of the electromagnetic field we need an inherently nonlinear theory. This requires finding a physical meaning for the negative energy states of the electron predicted by Dirac's equation. Dirac, himself, in his report of 1934~\cite{Dirac1934} to the \emph{7th council Solvay de Physique, Structure et Propri\'et\'es des Noyaux Atomiques}, entitled \emph{``Th\'eorie du positron''}, writes: \emph{``An electron in a state of negative energy is an object altogether foreign to our experience, but which we can nevertheless study from a theoretical point of view; \ldots{} Let us admit that in the universe as we know it, the negative energy states are almost all occupied by electrons, and that the distribution thus obtained is not accessible to our observation because of its uniformity throughout the space. Under these conditions, any state of unoccupied negative energy representing a rupture of this uniformity, must reveal itself to the observation as a kind of a hole. It is possible to admit that these vacancies constitute the positrons''}. An illustration of what Dirac described is shown is Fig.~\ref{fig:Diracsea}.

Dirac's quantum vacuum model cannot be completely consistent because the set of electrons in a vacuum have infinite charge density and current density. The model shows anyway that the phenomenon of pair creation can be interpreted as a transition from a vacuum electron to a positive energy state (electron) under the action of the electromagnetic field, which also creates a hole (positron) in the vacuum.

Calculations concerning the quantum vacuum are obviously complicated by this lack of consistency. To describe a real situation, we need to subtract one infinite amount (undisturbed vacuum energy) from another (vacuum energy disturbed, for example, by a magnetic field). Dirac~\cite{Dirac1934}, Euler~\cite{Euler1935}, Heisenberg~\cite{Heisenberg1936}, Weisskopf~\cite{Weisskopf1936} have found a method to do this subtraction which leaves no doubt. The method is based on the consideration that the energy, the current density, the charge density of the electrons of the vacuum, the electrical and magnetic polarisability of the vacuum, when no field is present, are quantities which correspond to divergent sums to which nevertheless one can attribute a clear physical meaning (see also~\cite{Schweber1994}). It is therefore possible to calculate the properties of the vacuum in the presence of electromagnetic fields and give the relations between $\mathbf{D}$, $\mathbf{H}$ and $\mathbf{E}$, $\mathbf{B}$. The electromagnetic characteristics of a vacuum can be described by an electrical permittivity and a magnetic permeability which depend on the fields present~\cite{Weisskopf1936}.

Indeed the mathematical form of the effective Lagrangian describing the interaction between the wave and the fields is fixed by the necessity that it must be a relativistic invariant. It therefore must be a combination of the relativistic invariants of the electromagnetic field $\left(\epsilon_0 E^2-B^2/\mu_0\right)$ and $\sqrt{\epsilon_0/\mu_0}\mathbf{E}\cdot\mathbf{B}$~\cite{Battesti2013}. One sees that for a plane wave which propagates in a vacuum since $E=B/c$ and $\mathbf{E}\cdot\mathbf{B}=0$ no non-linear effect is possible.

In the case of classical Maxwell electrodynamics, the principle of superposition holds. The electromagnetic fields associated to light just sum up with the external electromagnetic fields and no interaction between the two is possible. Maxwell's theory is intrinsically linear, the energy density of the field can be written as 
\begin{equation}
U_0 = \frac{1}{2}\left(\epsilon_0 E^2+\frac{B^2}{\mu_0}\right).
\end{equation}
Now, Weisskopf's predictions is that~\cite{Battesti2013}
\begin{align}\label{Ude}
\nonumber  U \approx &{1 \over 2}\left(\epsilon_0 E^2 + {B^2 \over \mu_{0}}\right) + {2\alpha^2 \hbar^3 \over 45 m_{e}^4 c^5} \left(\epsilon_0 E^2 - {B^2 \over \mu_{0}}\right)\left(3\epsilon_0 E^2 + {B^2 \over \mu_{0}}\right) \\
   &+{14\alpha^2 \hbar^3 \over 45 m_{e}^4 c^5}{\epsilon_{0} \over \mu_{0}} (\textbf{E} \cdot \textbf{B})^2.
\end{align}

The theory underwent a new development during the fifties when Feynman, Schwinger and Tomonaga formulated Quantum ElectroDynamics (QED), the success of which continues to this day. In 1949, Feynman presented a new positron theory~\cite{Feynman1949a}: \emph{``the problem of the behaviour of positrons and electrons in given external potentials, neglecting their mutual interaction, is analyzed by replacing the theory of holes by a reinterpretation of the solutions of the Dirac equation. \ldots{} In this solution, the ``negative energy states'' appear in a form which may be pictured \ldots{} in space-time as waves traveling away from the external potential backwards in time. Experimentally such a wave corresponds to a positron approaching the potential and annihilating the electron''}. We recognize in these words what we normally call Feynman's diagrams. In the appendix to the same article~\cite{Feynman1949a} we read: \emph{``a proof of the equivalence of the method to the theory of holes \ldots{} is given''} which implies that the results given by Heisenberg and Euler are once again obtained with the new method of Feynman.

Schwinger also later confirms the nonlinearity of the propagation of light in vacuum. A very detailed historical study of the QED birth can be found in~\cite{Schweber1994}. In his 1950 article~\cite{Schwinger1951}, we find: \emph{``Quantum Electrodynamics is characterized by several formal invariance properties, notably relativistic and gauge invariance. Yet specific calculations by conventional methods may yield results that violate these requirements, in consequence of the divergences inherent in present field theories \ldots{} Explicit solutions can be obtained in the two situations of constant fields, and fields propagated with the speed of light in the form of a plane wave. For constant (that is, slowly varying) fields, a renormalization of field strength and charge yields a modified Lagrange function differing from that of Maxwell field by terms that imply a non linear behaviour for the electromagnetic field. The result agrees precisely with one obtained some time ago by other methods and a somewhat different viewpoint''}, \emph{i.e.} the results of Heisenberg and Euler.

Nonlinear optical effects in a vacuum are thus very well founded on the basic principles of modern physics. 

A vacuum in quantum electrodynamics is therefore very different from the classical one which is simply nothing. The quantum vacuum has properties that electromagnetic fields can change. In 1951, Dirac sent a letter to the editors of Nature titled \emph{``Is there an {\AE}ther?''}~\cite{Dirac1951} in which he states : \emph{``We can now see that we may very well have an {\ae}ther, subject to quantum mechanics and conforming to relativity, provided we are willing to consider the perfect vacuum as an idealized state, not attainable in practice. From the experimental point of view, there does not seem to be any objection to this. We must make some profound alterations in our theoretical ideas of the vacuum. It is no longer a trivial state, but needs elaborate mathematics for its description.''} (see also~\cite{Navarro2018}).

Dirac failed to interest physicists to a new ether theory, but it is anyway true that the definition of a vacuum in quantum physics is complex. For example, Aitchison~\cite{Aitchison1985} gives the following: \emph{``The basic theoretical entity -- the quantum field -- is here regarded as analogous to a quantum-mechanical system with (infinitely) many degrees of freedom. A system of interacting quantum fields is then analogous to a complicated system in solid state physics; it can exist in different energy states, namely the ground state and various excited states. The excited states of the field system are characterized by the presence of excitation quanta, which are the particles (electrons, quarks, photons, ...) of which our material world is composed. In the ground state of the field system there are no excitation quanta, and hence no particles, present: the vacuum is  the ground state.''}

The sixties saw a new attempt to detect a variation in the refractive index $n$ of a vacuum in the presence of a magnetic field by R. V. Jones of the University of Aberdeen, Scotland~\cite{Jones1961} \emph{``In principle, the experimental arrangement has been to direct a beam of white light in vacuo through a `prism' of transverse magnetic field, similar to that used by Rabi (1929) in his improvement of the Stern-Gerlach experiment, and to detect photoelectrically any deflexion of the light beam by the field''}.

\begin{figure}
\centering
\includegraphics[width=0.8\linewidth]{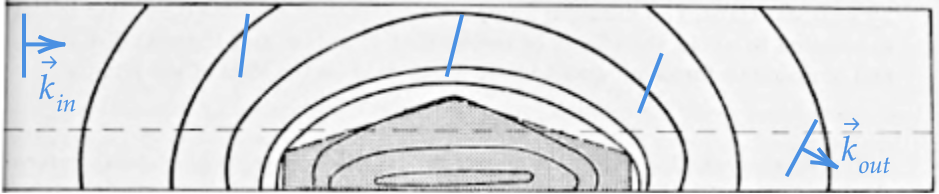}
\caption{Jones' experiment: a magnetic pole creates a strong magnetic field gradient, represented by contours of constant field strength; light wavefront propagating near it goes slower than the wavefront propagating farther from the pole. This results in a deviation of the plane wave. Our illustration of the deviated light path is exaggerated for the illustration purposes, the actual light path is shown in dashed line. Extracted from~\cite{Jones1961}.}
\label{fig:Jones Experiment}
\end{figure}

The polarised white light passed close to a magnetic pole of 0.8~T. The magnetic field gradient should induce a spatial variation of $n$ and a corresponding deflection of the light, see Fig.~\ref{fig:Jones Experiment}. The sensitivity of Jones' apparatus in refractive index  was of the order of $2\times 10^{-13}$.

The author of the article also claims that: \emph{``A tangible effect would be contrary to the principles of classical electromagnetism, and it appears \ldots{} that none has so far been predicted from quantum electrodynamics, if an effect were found, it would thus be of much interest. The experiment, while being unlikely to yield a positive result, is therefore among the class of fundamental null experiments which are worth repeating whenever a substantial gain in sensitivity is available''}.

The experimental method of Jones is quite original for a terrestrial laboratory, but the same phenomenon is quite common in the cosmos, in particular around pulsars. These are celestial objects that have magnetic field up to $10^9$~T. Electromagnetic radiation emitted from the pulsar or passing close to it cannot go in a straight line but it is obliged to follow field gradient like in Jones' experiment (\emph{e.g.} reference~\cite{Dupays2005} and references within). 

On the other hand, in 1961 T. Erber published a journal article in Nature on the \emph{``Velocity of Light in a Magnetic Field''}~\cite{Erber1961}. Erber presents the experiments already carried out and he proposes different other methods. Moreover, at the theoretical level, he makes the connection between the effect and the Lagrangian of Euler and Kochel~\cite{Euler1935}, also citing Weisskopf~\cite{Weisskopf1936}, as well as the connection between the Cotton-Mouton effect in standard media and the vacuum one. 

Actually, the effect of a magnetic field perpendicular to the propagation of light in a material medium has been discovered at the beginning of 20th century. Kerr in 1901 and Majorana in 1902 observed a weak birefringence created by a magnetic field transverse to the direction of light propagation in a suspension of Fe$_3$O$_4$ in water~\cite{Kerr1901} and in colloidal solutions of Fe~\cite{Majorana1902_0,Majorana1902_1}. This effect is also known as the Cotton-Mouton effect, since between 1905 and 1907, Cotton and Mouton have published at least 21 papers concerning this effect in liquids and colloidal solutions~\cite{CottonMemoire}. Linearly polarised light which propagates in a medium subjected to a magnetic field $\mathbf{B}$ perpendicular to the direction of propagation becomes elliptically polarised.

In other words, the speed $v_{\parallel}$ and the refractive index $n_{\parallel}$ for light polarised parallel to the magnetic field are different from the speed $v_{\perp}$ and the refractive index $n_{\perp}$ for light polarised perpendicular to the magnetic field. The difference $\Delta n$ between $n_{\parallel}$ and $n_{\perp}$ is proportional to $\textbf{B}^2$. In more general optical terms this is what is called a linear birefringence.

As a matter of fact, the calculation of $\Delta n$ was only done in the seventies~\cite{Bialynicka-Birula1970,Adler1971} giving the expectation value: 
\begin{equation}
\Delta n_{VCM} = {2\alpha^2 \hbar^3 \over 15 m_{e}^4 c^5}{B^2_0 \over \mu_{0}}\approx 4 \times 10^{-24}\, B^2_0
\end{equation}
where $B_0$ is given in Tesla units. Moreover, the effect does not depend on the light wavelength~\cite{Battesti2013}.

It is important to stress that an electric field $E_0$ would also give a similar linear birefringence
\begin{equation}
\Delta n_{VK} = {2\alpha^2 \hbar^3 \over 15 m_{e}^4 c^5}{\epsilon_0E^2_0}
\end{equation}
so that a 1~T magnetic field gives the same effect than an electric field of about 300~MV/m, the conversion factor being $c$. Experimentalists have therefore concentrated their efforts on $B$ because it is technically much simpler to produce a 1~T field than a 300~MV/m field.

Quantum electrodynamics therefore predicts for vacuum effects similar to what exist for material media, but how does this value compare with standard medium effects? Is it unmeasurably small?

\begin{figure}
\centering
\includegraphics[width=0.6\linewidth]{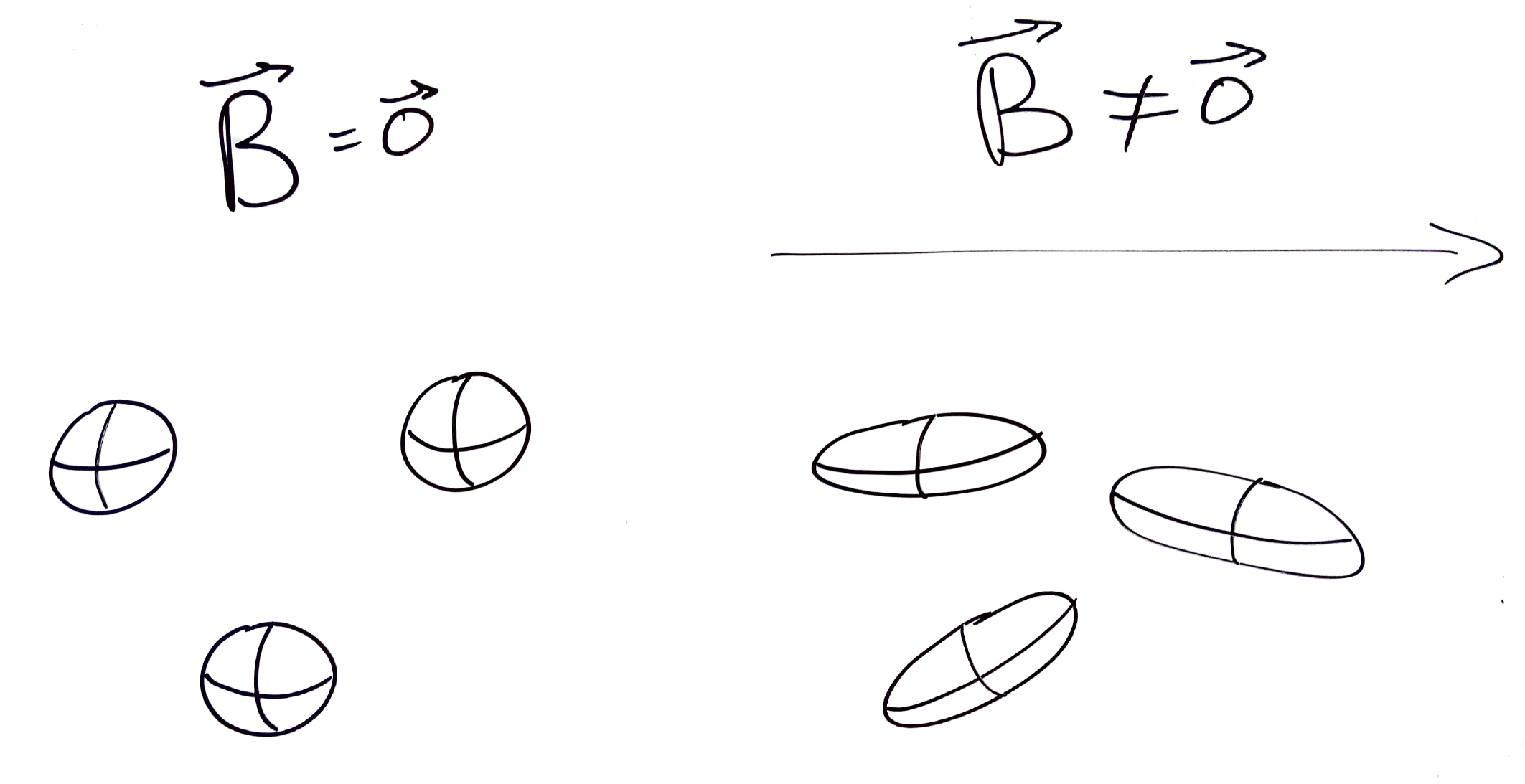}
\caption{Vacuum as a gaz: according to QED vacuum has a non vanishing Cotton-Mouton effect it thus act like a low pressure noble gas whose electronic cloud is deformed by the magnetic field.}
\label{fig:VacuumAsAGaz}
\end{figure}

The gases which show the weakest Cotton-Mouton effect are the rare ones of small atomic number: neon and helium~\cite{Rizzo1997}. For helium, the first measurement dates from 1991~\cite{Cameron1991a}; at a temperature of 20$^\circ$C, for a pressure of 1~bar and in the presence of a magnetic field of 1~T the expected value of $\Delta n$ is around $2.4\times 10^{-16}$. Under the same conditions, around $2\times 10^{-5}$~mbar, \textit{i.e.} a density of about $5\times 10^{17}$ helium atoms per m$^3$, give the same effect expected for a vacuum. From this point of view, vacuum ``atomic'' effect is macroscopic, so important that one can measure it even in the presence of a residual gas pressure of $2\times 10^{-9}$~mbar of O$_2$~\cite{Rizzo1997}. From this point of view, a vacuum acts therefore as a very subtle noble gas whose atomic current distribution is deformed by the presence of the magnetic field, see Fig.~\ref{fig:VacuumAsAGaz}.

\begin{figure}
\centering
\includegraphics[width=0.6\linewidth]{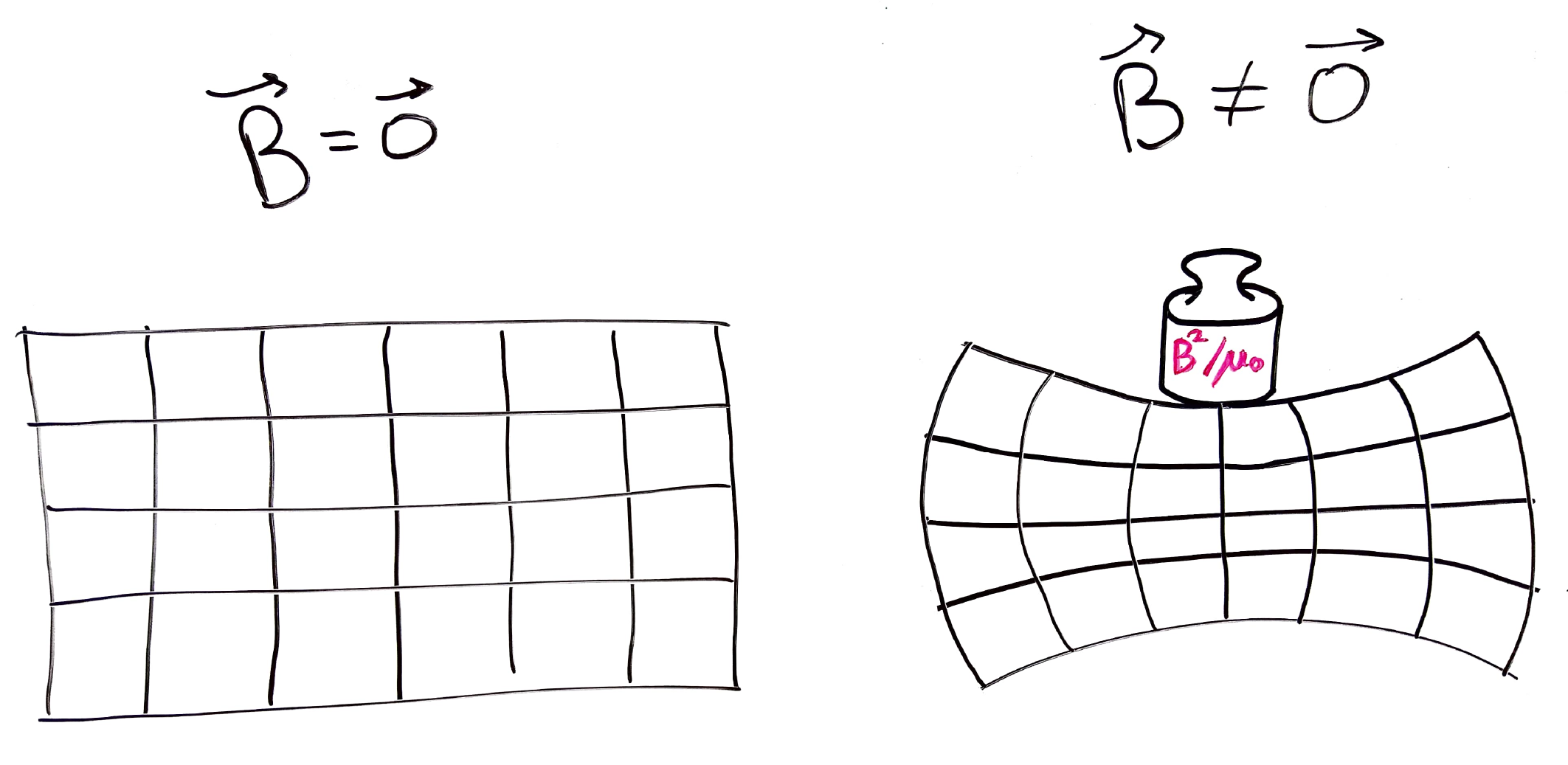}
\caption{Vacuum as a lattice: according to QED vacuum has a non vanishing stress birefringence. Being pressured along one of its axis by $B^2/\mu_0$ induce a birefringence.}
\label{fig:VacuumAsALattice}
\end{figure}

The analogy of a vacuum with a gas suffers from the fact that a gas is always associated with thermal agitation, a vacuum should be thought, maybe, as a crystal which is perfectly transparent, homogeneous and isotropic. Solid standard media under pressure become birefringent anyway. This is what is called the stress birefringence. Actually, $B^2/\mu_0$ has the dimensions of a pressure. For example, 10~T corresponds to about $8\times 10^6$~Pa. Vacuum $\Delta n$ can be therefore written as $\Delta n = C (B^2/\mu_0)$ with $C \approx 5\times 10^{-30}$~Pa$^{-1}$. For the sake of argument, let's compare with BK7 glass. For this material $C$ is around $2.7\times 10^{-12}$~Pa$^{-1}$~\cite{SOM}. From this point of view, a vacuum behaves as it has a lattice that is disturbed by the magnetic field pressure, see Fig.~\ref{fig:VacuumAsALattice}.

A gas or a crystal are imaginative analogies but one has to be very careful not to let the ether show up again in physics.

\section{Contemporary times}

The publication of the $\Delta n$ value marks the beginning of an important and continuous experimental activity together with the publication of several proposals of experiments. Many reviews have therefore been published in recent and less recent years detailing the theoretical aspects, the different experimental methods, and results from an historical and technical point of view (see \emph{e.g.}~\cite{Battesti2013,HIMAFUN}), information about astrophysical tests can be found as well.

The main event was in 1979 when E. Iacopini and E. Zavattini published a project~\cite{Iacopini1979} which was finalized in 1993~\cite{Cameron1993} thanks to a collaboration with A. Melissinos. The experiments which are currently in development~\cite{BMV,OVAL} are different versions of this original idea and the experiment PVLAS that has reported the best limits on vacuum magnetic birefringence was also originated by Zavattini~\cite{PVLAS}.

\begin{figure}
\centering
\includegraphics[width=0.9\linewidth]{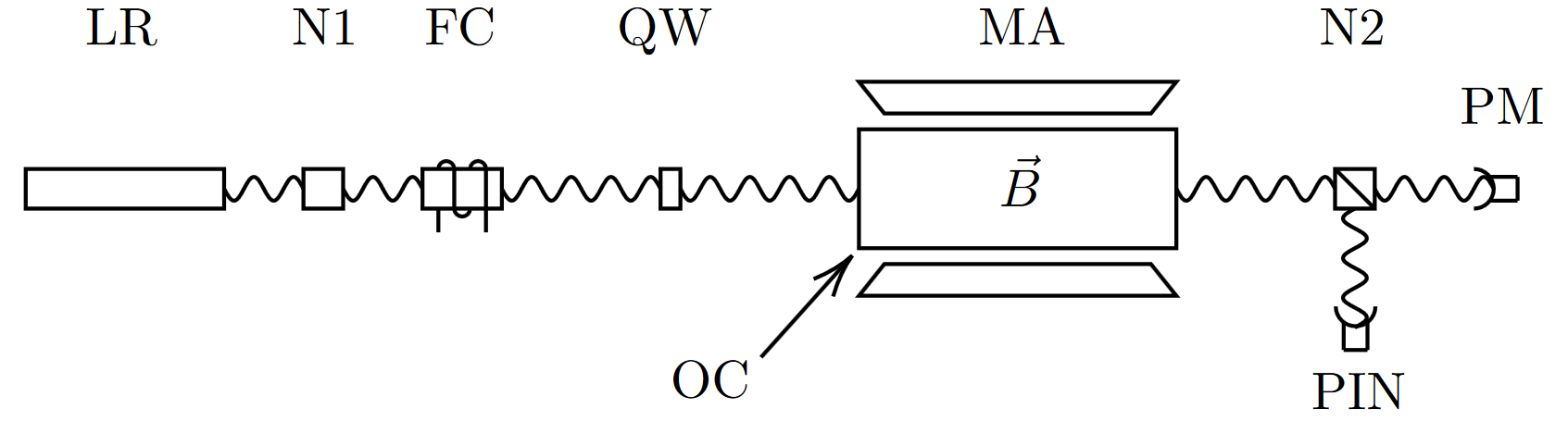}
\caption{Iacopini and Zavattini proposal to measure the vacuum magnetic birefringence using a polarimetry technique. The light from a laser, LR, is polarised by a first polariser, N1, the polarisation is then modulated by a ellipticity modulator -- the combination of a Faraday cell FC and a quarter wave plate QW -- an optical cavity, OC, is used to increase the signal induced by the transverse magnetic field $\vec{B}$ provided by the magnet MA. Finally the exiting polarisation is analysed using a second polariser, N2, and two photodiodes, PIN and PM.}
\label{fig:ZavattiniProposal}
\end{figure}

The experimental design proposed in 1979~\cite{Iacopini1979} is an improved version of the device initially used by Buckingham, Prichard and Whiffen~\cite{Buckingham1967} and then by different teams for the measurements of the Cotton-Mouton effect in gas~\cite{Rizzo1997}, it is represented in Fig.~\ref{fig:ZavattiniProposal}.

One gives up the direct measurement of $v_\parallel$ ($n_\parallel$) or $v_\perp$ ($n_\perp$), measuring the anisotropy of the refractive index $\Delta n$ by measuring the ellipticity $\Psi$ acquired by the light of wavelength $\lambda$ by propagating along an optical path of length $L$ in the presence of a magnetic field $\mathbf{B}$
\begin{equation}
\Psi=\pi\frac{L}{\lambda}\Delta n.
\end{equation}

The light, before passing in the magnetic field region, passes through an optical system which modulates ellipticity at a carrier frequency $\nu_{ca}$. The polarisation is finally analysed by a polarising prism crossed at 90$^\circ$ to the initial polarisation and the light is detected by a photodiode. It can be shown~\cite{Rizzo1997} that the signal from the photodiode contains a modulation at the frequency $\nu_{ca}$ with an amplitude proportional to $\Psi$ (homodyne technique). This assembly is essentially limited by the static birefringences always present in an optical system~\cite{Rizzo1997}. With a device of this type, H\"uttner's team was able to arrive at a sensitivity, on the measure of $\Delta n$, of $2.5\times 10^{-16}$ with a $B^2$ of 1~T$^2$~\cite{Huttner1987}.

The main innovation  of the 1979 proposal~\cite{Iacopini1979} consists in introducing a modulation at the frequency $\nu_{m}$ of the ellipticity $\Psi$ to be measured, by modulating the supply current of the magnet which gives the field $\mathbf{B}$~\cite{Cameron1993}. This is a technique called heterodyne, the photodiode signal contains modulations at frequencies $\nu_\pm = \nu_{ca} \pm \nu_{m}$~\cite{Rizzo1997}. The optical path in the region subjected to the magnetic field is also increased by using optical cavities which are non-resonant in the original version of the experiment~\cite{Iacopini1979,Cameron1993} and resonant in the PVLAS version~\cite{PVLAS}. The sensitivity of the device was thus improved and it was possible to measure for the first time the Cotton-Mouton effect of neon~\cite{Cameron1991b} and helium~\cite{Cameron1991a} thanks to an experiment carried out at Brookhaven National Laboratories, NY (USA)~\cite{Cameron1993}. The value of $B^2$ was about 4~T$^2$ and the optical path was more than one kilometer, the results were still consistent with zero; the experimentalists nevertheless arrived at ``only'' five orders of magnitude of the Cotton-Mouton effect predicted by quantum electrodynamics.

\begin{figure}
\centering
\includegraphics[width=0.6\linewidth]{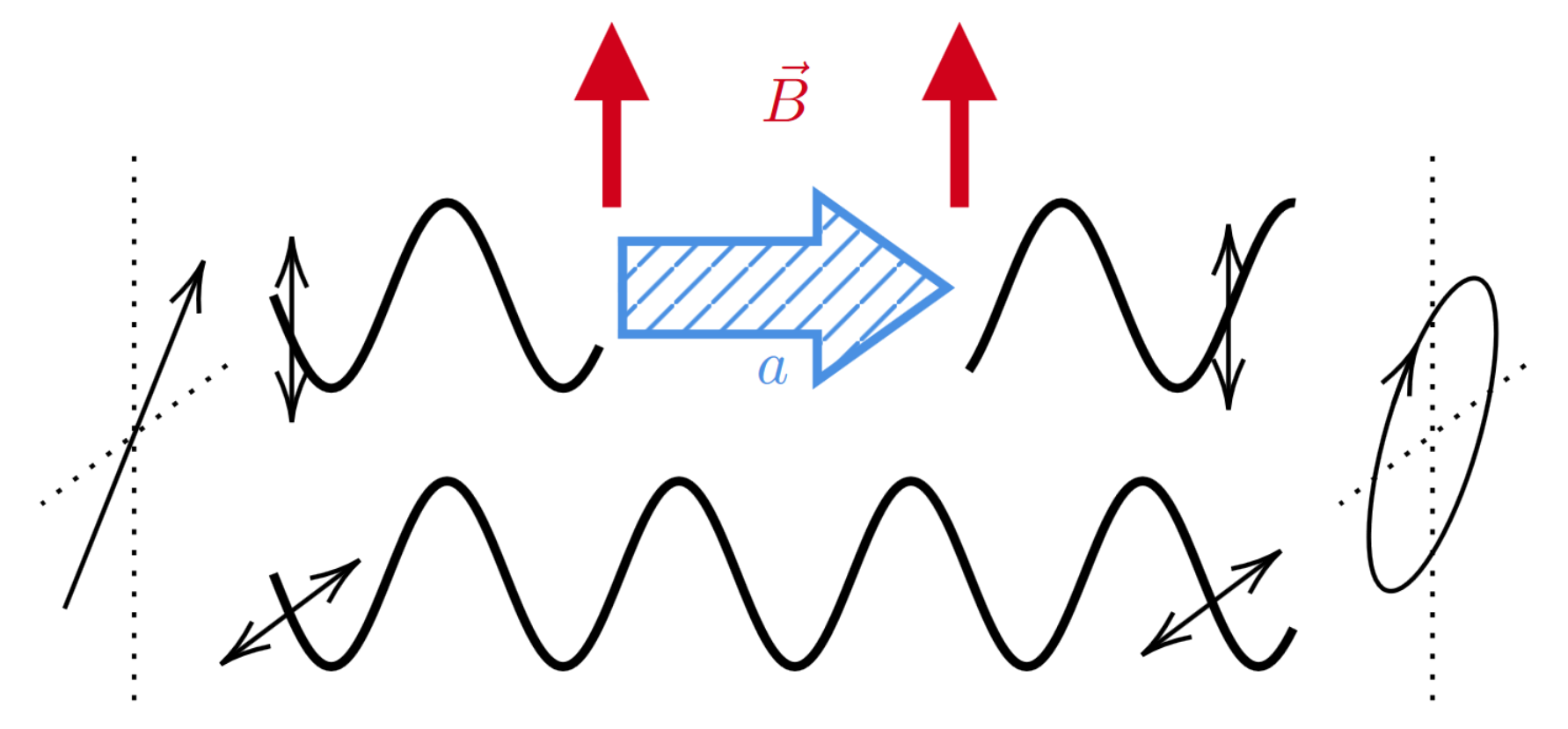}
\caption{Hypothetical axion particles can contribute to the vacuum magnetic birefringence. Indeed, the transverse magnetic field can produce axions following the light polarisation. Because axions are massive, they travel slower than light, and when becoming a photon back they induced some phase shift in only one polarisation. This corresponds to a birefringence.}
\label{fig:Axion}
\end{figure}

Actually, the Brookhaven experiment was not only motivated by the QED effect but also by the fact that light speed could be affected by the presence of a magnetic field because of the existence of particles predicted beyond the Standard Model. As first reported in~\cite{Maiani1986}, particles like the axions could couple with photons via the magnetic field. A photon during its propagation in the magnetic field could become an axion-like particle (ALP) propagating at a lower velocity since ALP can be massive, and then interacting with the magnetic field becoming a photon again. Only photons polarised parallel to the magnetic field can oscillate into ALPs, therefore $n_\parallel \neq n_\perp$ which corresponds to a birefringence, see Fig.~\ref{fig:Axion}.

\begin{figure}
\centering
\includegraphics[width=0.5\linewidth]{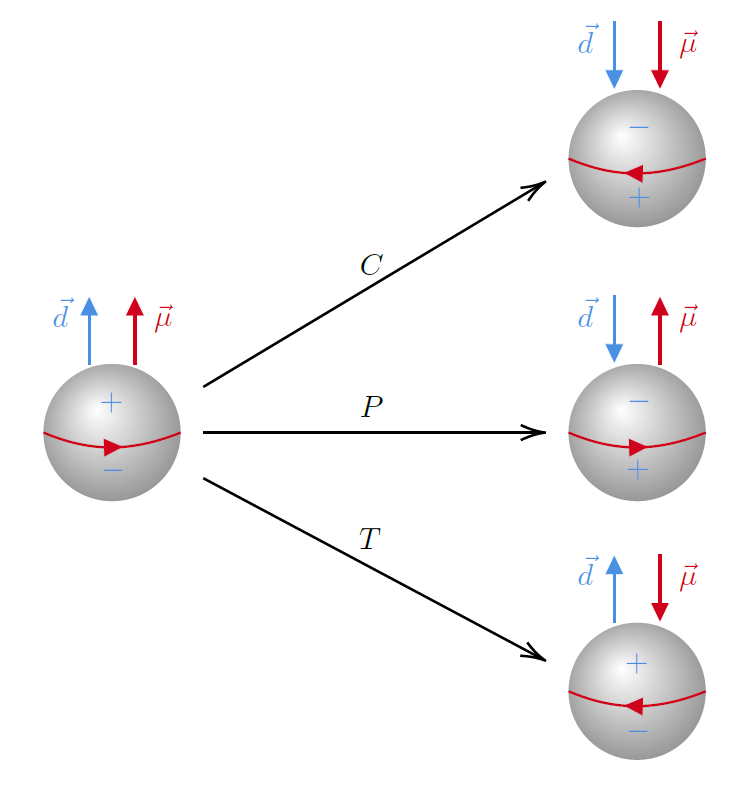}
\caption{Representation of the symmetry transformation P and T on a neutron with a magnetic and electric dipole $\bm\mu$ and $\mathbf{d}$. The Standard Model predicts a CP symmetry violation therefore a non vanishing electric dipole for the neutron is predicted as well. But, experiments give zero results: this is the strong CP problem.}
\label{fig:neutronDipole}
\end{figure}

Axions have been introduced to solve what is called the strong CP problem~\cite{Peccei1977} \emph{i.e.} why does quantum chromodynamics seem to preserve CP symmetry? This problem manifests itself in an experimental zero neutron electric dipole moment when theory predicts a non zero value unless a very unlikely fine tuning of some theoretical parameters. Fig.~\ref{fig:neutronDipole} shows how an electric dipole of the neutron breaks the CP symmetry.

The end of the Brookhaven experiment was accompanied by the beginning of a novel experiment, PVLAS, whose final results have been published in 2020 after more than 25 years of work~\cite{PVLAS}. The limit reported is the best ever reached: $\Delta n \leq 2\times 10^{-23}$ for a 1~T field. The PVLAS result is still compatible with zero but at a level that clearly indicates that the QED prediction for vacuum magnetic birefringence is not unmeasurably small.  One of the particularities of this experiment is the use of DC magnets, superconductive or permanent, and rotating them to modulate the effect instead of modulating the magnet driving current like in the Brookhaven one. Rotating the magnet corresponds to rotating the birefringence axis and therefore modulating the ellipticity to be measured.

\begin{figure}
\centering
\includegraphics[width=0.6\linewidth]{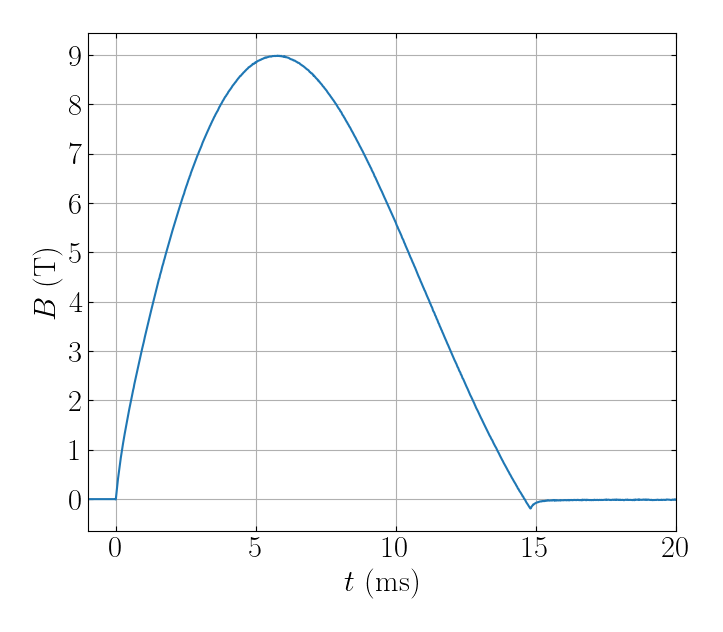}
\caption{Typical time dependence of a pulsed field of the BMV experiment, the time of rise is about 6~ms with a $B_{max}$ as high as 9~T.}
\label{fig:PulsedField}
\end{figure}

Nowadays, notwithstanding that many experimental proposals have been put forward~\cite{HIMAFUN}, only two experiments are under way~\cite{BMV,OVAL}. Both are based on the use of pulsed fields, as first suggested in~\cite{Rizzo1998}, a technique that in principle allows to reach fast modulated fields higher than 10~T, see Fig.~\ref{fig:PulsedField}.

This has certainly some advantages but also creates a certain number of technical problems as discussed in~\cite{RSIBMV2021} that have not yet permitted to reach the PVLAS level using pulsed fields.

\section{Conclusions}

So, does the speed of light in a vacuum change in the presence of a magnetic field?

Following QED the answer is yes. It changes of about a few parts in $10^{24}$ in a 1~T field depending on the polarisation of light. Unfortunately we have not yet been able to prove it experimentally. QED predictions have been proved to be correct very precisely for charged particles and bound systems (see~\cite{Sailer2022} and references within). In all these kind of experiments, electromagnetic waves and its constituent particle the photon are typically used as probes, for example to measure the QED corrections to an energy level. Here, the behaviour of electromagnetic fields themselves is modified following QED, and moreover, while all other phenomena are microscopic, here the predicted variation of light velocity is one of the rare macroscopic manifestation of QED. If successful, in an experiment like PVLAS the expected ellipticity would have been accumulated thanks to about 500~km of propagation in the magnetic field. Conceptually, it is very different to say that in order to write correctly the mathematical form of the electric dipole field of an electron, one has to add QED corrections to the classical expression (\emph{e.g.}~\cite{Fouche2016}), and to say that a vacuum acts exactly like a material medium on a macroscopic level. In a sense, QED is based on special relativity, which is supposed to get rid of the ether idea, but it resulted in a new form of the same ether concept with in addition a sensitivity to the presence of electromagnetic fields exactly like, for example, a macroscopic quantity of Helium gas. This is quite puzzling. 

Since QED predictions seem to be always correct for charged particles and bound systems, we expect the same for electromagnetic waves, but in a quantum vacuum everything exists in a virtual state, what we know and what we do not know yet. The axion particles are the example that we have talked about in the previous section where we have recalled that the propagation of light in a vacuum can be affected by their virtual existence in the presence of a magnetic field. Experimental results could therefore open a new window on the cosmos, changing once again what we use to describe it and giving the opportunity to future physicists to add several new pages to the history we reviewed in this paper.

\bibliography{BibEPJH}

\end{document}